# 2D MoS$_2$ under switching field conditions: study of high-frequency noise from velocity fluctuations

J. M. Iglesias*
*Department of Applied Mathematics, University of Salamanca, Salamanca 37008, Spain.*

E. Pascual, S. García-Sánchez, and R. Rengel
*Department of Applied Physics, University of Salamanca, Salamanca 37008, Spain.*
(*josem88@usal.es.)
(Dated: July 4, 2023)

The transient high-frequency noise response of two-dimensional MoS$_2$ under abrupt large signal switching field conditions is studied by means of an ensemble Monte Carlo simulator. Low-to-high and high-to-low transitions are analyzed at low (77 K) and room temperature, considering several underlying substrates. The incorporation of stochastic individual scattering events allows capturing the transient collective phonon-electron coupling, which is shown to be responsible for the appearance of an oscillatory behaviour in the average velocity and energy at low temperature in the case of MoS$_2$ on SiO$_2$, hBN and Al$_2$O$_3$. Activation and deactivation of surface polar phonon emissions in the low-to-high field switching process yield to the appearance of a relevant peak in the power spectral density of velocity fluctuations in the THz range. The results show the important influence of the substrate type in the noise behaviour of MoS$_2$ at very high frequencies, which is critical for the design of future FET devices based on 2D TMD technology.

Transition metal dichalcogenides (TMDs) are becoming increasingly relevant due to their unique electronic, mechanical, and optical properties. In their monolayer version, they feature a very thin and flexible atomic layer structure which gives them distinctive mechanical and optical properties. Additionally, they exhibit a direct bandgap well over 1 eV, making them excellent candidates for high-speed, low-power electronic devices such as transistors and integrated circuits. TMDs are also promising for the development of emerging technologies applied to different electronic and optoelectronic devices [1]. In particular, their reliability as multifunctional diodes[2], transistors[3, 4], photodetectors[5, 6], solar cells[7, 8], flexible electronics[9, 10], or biosensors[11] has been demonstrated.

The investigation of noise in TMDs technologies is critical for the development of future commercial applications. Up to date, noise in TMDs (and in particular, MoS$_2$, the most relevant material within this category) has been studied from two main perspectives. First, from a circuital point of view, where the efforts have been mostly related to the determination of the noise margins in TMD-based CMOS inverters: a recent study has found a nearly ideal noise margin in inverters combining WSe$_2$ and MoS$_2$ FETs [12]. Second, from a device perspective, where several studies can be found related to the evaluation of low-frequency noise in MoS$_2$-FET devices, which is critical for signal sensing and processing applications [13]. It has been found that 1/f noise in single layer MoS$_2$ is due to fluctuations in carrier mobility, with the Hooge parameter behaving similarly as in organic thin-film transistors and graphene FETs [14], while in multiple layer MoS$_2$ 1/f noise has been attributed to carrier number fluctuations [15]. Finally, the study of the noise equivalent power in MoS$_2$ photodetectors has demonstrated the extremely sensitive detection capability of these devices [16]. However, due to the newness of these materials, there is still a significant lack of studies relative to high-frequency noise, with particular focus on the origin and importance of intrinsic noise sources in MoS$_2$. The Monte Carlo modelling technique has proven to be very useful for this analysis due to its numerical and stochastic nature, which allows facing the investigation of high-frequency noise in a natural way [17]. On the other hand, studying transient regimes under large signal-switching field conditions is relevant for both analog and digital transistor applications. In particular, having detailed knowledge about non-stationary phenomena related to transient velocity fluctuations and their impact on noise is important in order to help designing reliable future TMD channel transistors with fast-switching capabilities.

In this work, we analyze the noise generated by instantaneous velocity fluctuations in MoS$_2$ (the most prominent TMD) during the transient regime that takes place after abrupt switching electric field conditions (both from low-to-high and high-to-low values) using an in-house ensemble Monte Carlo (EMC) material simulator[18, 19]. The simulations are performed in the reciprocal space, which is equivalent to considering an infinitely large 2D material sheet, thus avoiding any issue relative to edge effects or modifications of the band structure due to small size constrictions. This modelling technique allows providing a microscopic insight into the intrinsic sources of noise in the material. Furthermore, thanks to this stochastic approach, individual scattering activity or quantities directly related to fluctuations (e.g., correlation functions or power spectral densities) are also accessible for their analysis, providing an enriched physical insight of the ultimate sources of high-frequency noise in the material. The physical model that we have considered assumes a parabolic ε-k dispersion relation for the description of the band structure, including K (primary) and Q (secondary) valleys in the first Brillouin zone [20]. For an accurate treatment of electronic transport, it is necessary to take into account the upper Q valleys, especially at high (ambient) temperatures and at the relatively high electric fields that are reached in the present work. The scattering probability is described by means of the deformation potential formalism, taking into account intra- and inter-valley acoustical phonon branches, optical phonon branches, scattering with surface polar optical phonons (SPPs) from the substrate, degeneracy, and the screening of free carriers [18, 19]. Undesired impurities and defects, which are known to be a notorious source of degradation in experimental samples due to the relatively immature degree of development of TMD fabrication technology, are not considered in this work in order to provide the intrinsic noise performance of $MoS_2$ in a best-case scenario.

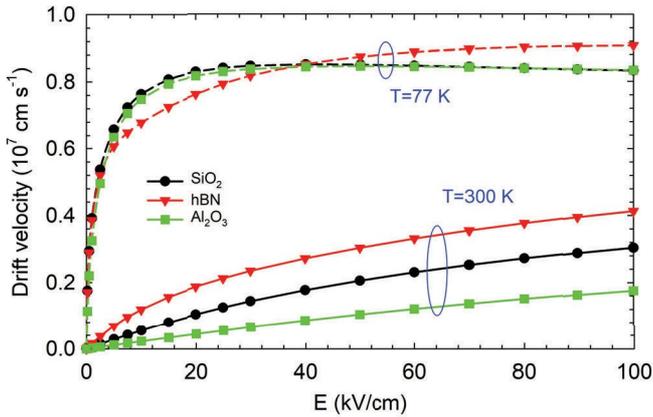

FIG. 1. Average carrier drift velocity as a function of the applied electric field at two temperatures (77 and 300 K) for $MoS_2$ standing on three different substrates, $SiO_2$, hBN and $Al_2O_3$.

In order to properly account for the possible influence of degeneracy on the velocity fluctuations results, we implemented the method proposed by Thobel et al [21] for the case of a two-dimensional electron gas, based on the consideration of an excess population governed by a linearized Boltzmann transport equation. The methodology was successfully used for the study of 2D materials (graphene) in [22]. It is important to mention that in the conditions in the present work (transient under large signal electric field-switching), the results obtained with excess and background populations are nearly identical at low and room temperature, thus indicating that degeneracy does not play a role in the transient evolution.

An exhaustive description of the physical model and the simulation framework can be found in the supplementary material.

Before analysing the switching regime, let us examine the low and high field regimes from a stationary perspective. Figure 1 shows the average drift velocity in $MoS_2$ as a function of the applied electric field up to 100 kV/cm at two temperatures, 77 and 300 K. Three different cases related to the underlying substrate are presented: $SiO_2$ (the most common substrate in practical applications), hexagonal boron nitride or hBN and $Al_2O_3$. At 77 K, there is a rapid increase of the average drift velocity (with larger values in the $SiO_2$ and $Al_2O_3$ cases), followed by a saturation that takes place approximately at 25 kV/cm. The saturation in $MoS_2$ on hBN occurs at larger fields, which leads to slightly larger velocity values from 40 kV/cm and beyond. At 300 K, the velocity is much lower for the studied cases, with larger values for $MoS_2$ on hBN in the whole electric field range. Moreover, there is a monotonic increase of the drift velocity, with no strict saturation. In the case of $MoS_2$ on $HfO_2$ (not shown in the graphs; the results can be examined in the supplementary information) the velocity reaches much lower values than in the case of considering other substrates, yielding to a severe degradation of the drift velocity and the low field mobility due to the extremely active SPP scattering in that case.

From the results observed in Fig. 1, we shall consider as low field condition an electric field value equal to 1 kV/cm, while for the high field value we have set the electric field to be 50 kV/cm. In this way, we may capture the relevant features of both regimes at both temperatures.

Let us examine the time transient regimes from low-to-high and high-to-low switching field. Figure 2 shows the time evolution of the average carrier drift velocity and energy at 77 K and 300 K, and for the $SiO_2$, hBN and $Al_2O_3$ substrates. The average transient velocity in the direction of the applied field presents a remarkable oscillatory effect at 77 K in the low-to-high field transient. The period of these oscillations depends on the type of substrate (0.13 ps for $MoS_2$ on $SiO_2$ and $Al_2O_3$ and 0.18 ps for $MoS_2$ on hBN, approximately), finally reaching a stationary state. As it will be discussed afterwards, this is related to the surface polar phonon scattering activity, which depends on the substrate type. On the other hand, the oscillatory effect is not present at 300 K, although a velocity overshoot is noticeable. At room temperature the stationary state is quickly reached, presenting a much faster transient than at 77 K. The curves at 300K show a noisier behaviour, which is the signature of instantaneous fluctuations directly associated with an increased scattering activity at room temperature.

In the high-to-low field transient, the above-mentioned velocity and energy oscillations in the transient are not present. However, an undershoot transient velocity takes place for $MoS_2$ on hBN, while for $MoS_2$ on $SiO_2$ and $Al_2O_3$ a monotonic decay is observed. On the other hand, the average energy presents an exponential-like decay, with the stationary state apparently reached after 2 ps. In general, larger energy values are obtained in the whole time range for $MoS_2$ on hBN in both the high-to-low and low-to-high cases, which directly derives from a weaker electron-SPP coupling in comparison to $MoS_2$ on $SiO_2$ and $Al_2O_3$. Since $SiO_2$ and $Al_2O_3$ provide a similar behavior, from now on the results will be presented only for $MoS_2$ on hBN and $MoS_2$ on $SiO_2$. The subsequent results for $MoS_2$ on $Al_2O_2$ and $MoS_2$ on $HfO_2$ are shown in the supplementary material for the interested reader.

In order to discuss the physical origin of the transient velocity and energy oscillations observed at low temperature in the low-to-high switching, in Figure 3 it is shown the time evolution at 77K of the inverse momentum relaxation time associated to the most relevant scattering mechanisms (intrinsic $MoS_2$ phonons and emissions from the lowest energy surface polar phonon from the substrate). As it can be observed, in the case of $MoS_2$ on $SiO_2$ after the abrupt field switching there is a very relevant increase in the number of scattering events related to surface polar phonons from the underlying substrate. In particular, these are related to the lowest energy phonon, and they correspond to SPP emissions in the primary K valleys. After reaching a maximum, oscillations with dimmed amplitude follow until a stationary state is reached. Such behaviour is explained by the following: the abrupt onset of a high field accelerates the carrier ensemble. As the carriers gain energy from the electric field, inelastic phonon emissions come progressively into play when individual electrons reach sufficiently high energies. Since SPPs from SiO$_2$ present a stronger electron-phonon coupling, they become dominant, as compared to intrinsic phonons, in the first instants after the field switching. These emissions with a dominant phonon imply an abrupt energy reduction of the electrons involved in such interactions. Therefore, the SPP scattering activity rapidly raises, but after the carriers energy drops below the phonon emission threshold, the scattering probability abruptly drops, allowing the carriers to gain energy collectively again. In this process of adaptation to the final stationary situation, there is a back-and-forth process related to SPP phonon emissions, which becomes progressively more appeased as the final carrier distribution is reached, as the rest of scattering events (with different phonon energy values) enter into play, and a collective velocity and energy oscillation is no longer present. SPP phonons related to hBN have larger energies and a reduced electron-phonon coupling. Consequently they are not the dominant scattering mechanism. This has an important effect as compared to the case of MoS$_2$ on SiO$_2$: in MoS$_2$ on hBN there is an important velocity overshoot, but the oscillations are rapidly damped. At 300 K (not shown in the graphs) the scattering activity increases very significantly for both substrates. This implies that the electron population is rapidly redistributed and the collective oscillatory behavior is no longer present. Consequently, the temperature influence can be considered to be dominant in order to observe or not the collective oscillatory effect. However, it is interesting to focus on the substrate influence at low temperatures, since these latter are required in order to achieve the best electronic properties in this material and provide a competitive carrier mobility as compared to conventional semiconductors.

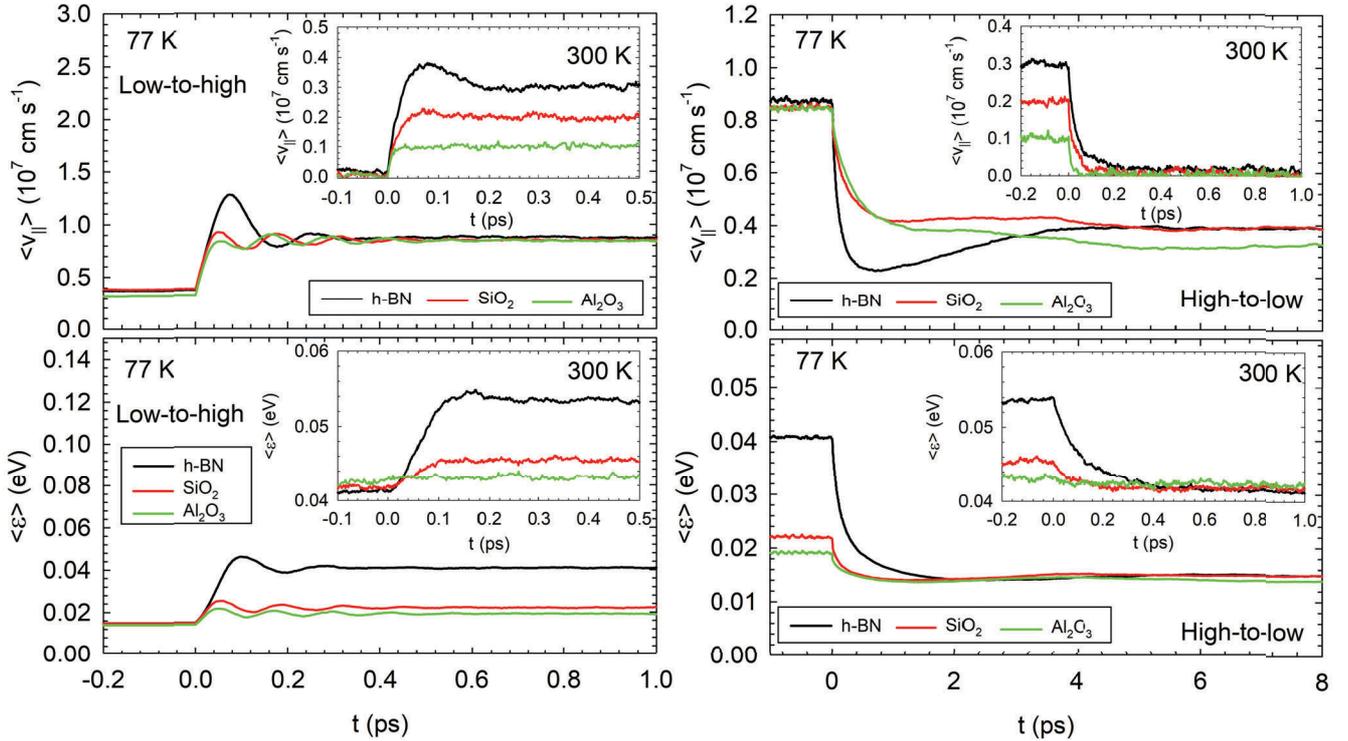

FIG. 2. Transient response of the average carrier drift velocity (in the direction of the applied field) and energy when switching from low to high fields (left) and high to low (right), at 77 K. Insets shows the results at room temperature.

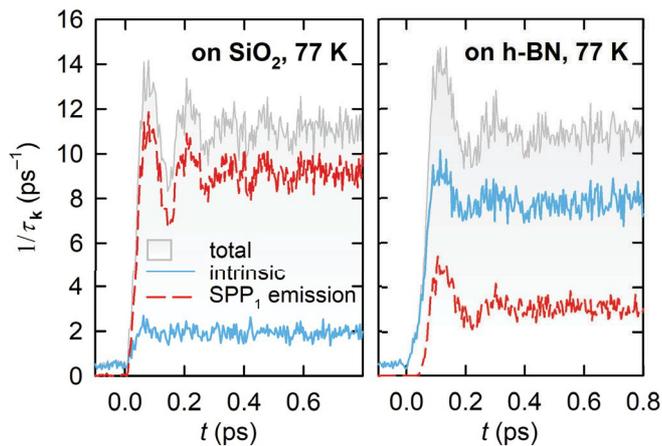

FIG. 3. Instantaneous momentum relaxation time in low to high field switch at 77 K in SiO$_2$ (left), and hBN (right) for the intrinsic phonon modes, smallest energy SPP emission, and the total. The shown SPP interaction accounts for almost all the electron-substrate phonon interactions.

The phonon-carrier interplay and the collective response not only affects average macroscopic quantities, but it has also an important impact on noise. The transient evolution of the velocity is accompanied by intrinsic instantaneous fluctuations, both determining the global noise behavior along the switching process. The process and numerical method employed to carry out the subsequent analysis of fluctuations from the EMC simulation is explained in the supplementary material and in prior works[23–26]. Figure 4 shows the two-time correlation function of velocity fluctuations ($C_{\delta v}(t, \tau)$) as a function of time and at different moments along the transient.

Let us first focus on the low-to-high switching. At 77 K, there is an oscillatory behaviour in $C_{\delta v}$, which is more pronounced in the MoS$_2$ on SiO$_2$ case. This is in direct relation to the collective phonon-carrier coupling discussed when the scattering events along the transient were presented, since the onset and deactivation of inelastic scatterings is a vector for breaking of velocity correlation and the change in fluctuation sign (as the individual velocity passes from being larger than the average to being smaller, or vice versa). The period of the oscillations in MoS$_2$ on hBN is larger and the amplitude is damped more rapidly. At 300 K, the oscillations in $C_{\delta v}$ disappear for SiO$_2$ substrate and are significantly appeased for hBN, and the characteristic times are much shorter, which is due to the strongly increased scattering activity at room temperature. In the high-to-low switching $C_{\delta v}$ shows a monotonic decay and oscillations are not present, with slightly larger correlation times.

Figure 5 shows the power spectral density of velocity fluctuations for the different cases above-mentioned, at different times in the transient switching process. In general, a progressive increase of the low-frequency value of the power spectral density is observed as the transient evolves, in correspondence with a larger zero-time value of the correlation function. At 77 K, a Lorentzian shape is observed along the whole transient, with a peak that becomes more pronounced as the final stationary state approaches. The peak is located at 7.6 THz for MoS$_2$ on SiO$_2$ and 5.5 THz for MoS$_2$ on hBN, approximately, which is in good accordance with the inverse period of the oscillatory behaviour of the correlation and its relation with the alternating activation of the substrate polar phonon with lower energy. Consequently, at 300 K, as it could be expected the peak is much less pronounced for the hBN substrate and hardly noticeable for the SiO$_2$ case. In the high-to-low case, the presence of peaks only takes place at 77 K, and its physical origin is the undershoot velocity effect which is responsible for the negative values of the correlation function. At 300 K, the maximum disappears due to the monotonic decay of the correlation function, since the correlation loss is dominated by a faster scattering activity as compared to the low-temperature case. As compared to other materials, such as InP and GaAs [25] or graphene[26], MoS$_2$ presents a much noisier behaviour evidenced by larger values of the power spectral density, their low-frequency values being approximately one order of magnitude larger than in those materials for low-to-high field transitions. In general, MoS$_2$ shows transient response times comparable to those of graphene [26]. It is interesting to notice that the peak values appearing in the low-to-high field transients and at the final stationary high-field states, which have been also observed in InP, GaAs and graphene, take place at significantly higher frequencies than in the case of these other materials (approximately 0.7 THz, 1.2 and 2.5 THz for InP, GaAs and graphene, respectively [25, 26]). Further analysis of the transient power spectral density of MoS$_2$ on HfO$_2$ and Al$_2$O$_3$ is included in the supplementary material.

To summarize, the intrinsic transient behaviour of noise related to velocity fluctuations under large signal switching conditions has been studied in MoS$_2$. At low temperature, carrier-phonon scattering events with the lowest energy SPP are critical to explain the transient behaviour of the material and the appearance of peaks in the power spectral density. In particular, the periodic activation and deactivation of the scattering related to that phonon (that depends on the substrate type) explains the physical origin of the THz behaviour of the noise related to the intrinsic velocity fluctuations. High-frequency noise in MoS$_2$ is therefore importantly influenced by the substrate type considered, specially at low temperatures (necessary to reach competitive velocity and mobility values), so an adequate selection of the underlying dielectric shall be critical in the design of low-noise future FET transistors based on a 2D TMD channel.

Supplementary material

The supplementary document that comes along with this work includes: (1) the details of the physical simulation model; (2) the numerical approach for the analysis of velocity fluctuations, and (3) further complementary results.

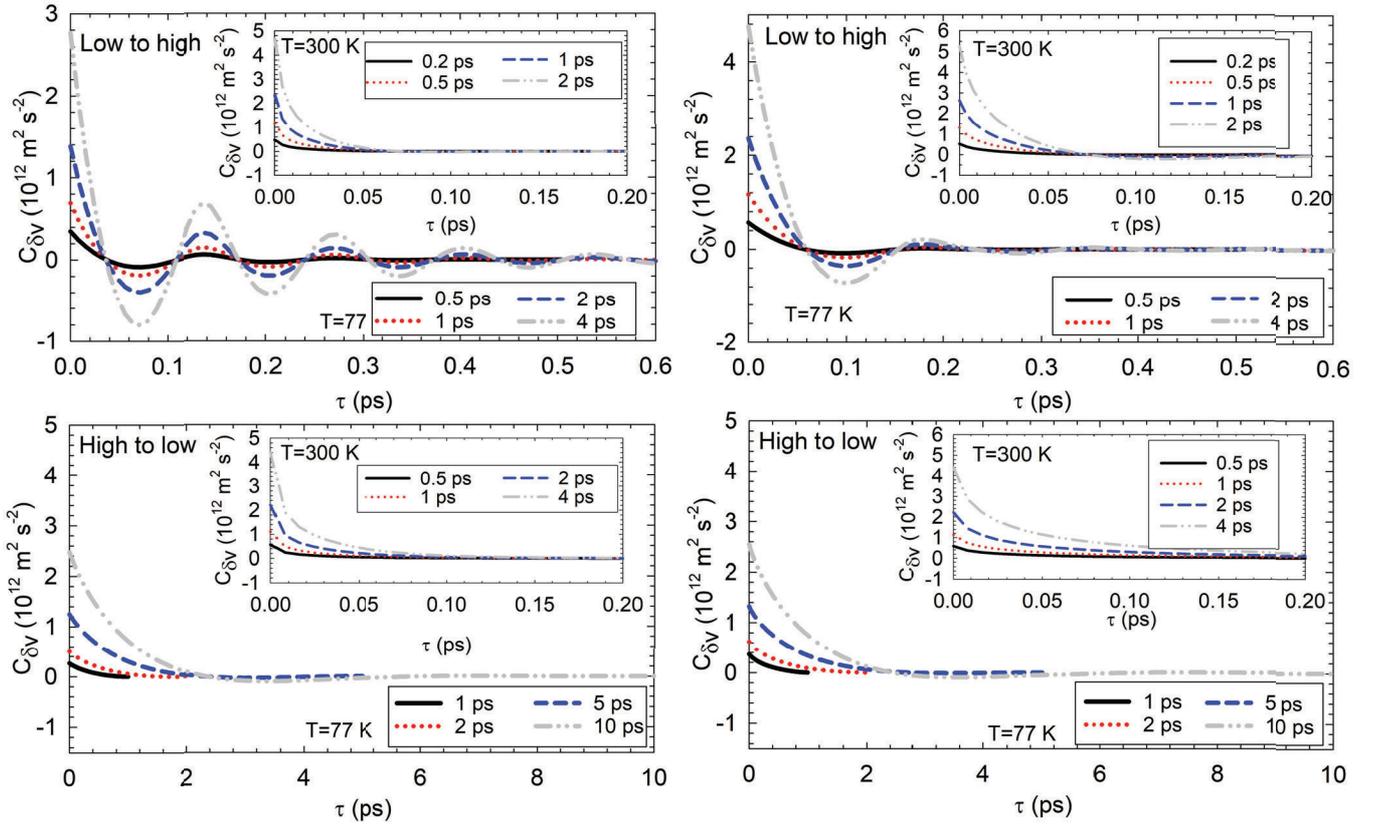

FIG. 4. Temporal correlation function of velocity fluctuations at different instants in the transient process for MoS$_2$ on SiO$_2$ (left) and MoS$_2$ on hBN (right), at 77 K. Insets shows the results at 300 K.

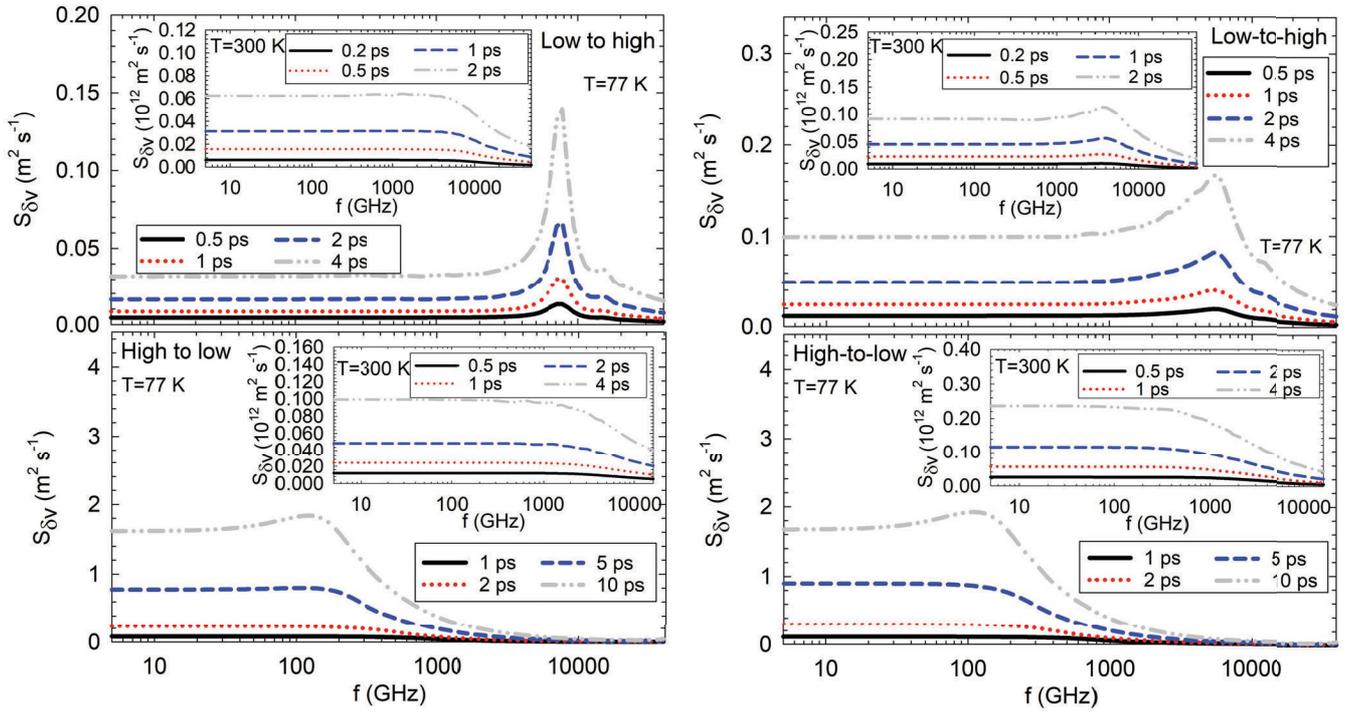

FIG. 5. Power spectral density of velocity fluctuations at different instants in the transient process for MoS$_2$ on SiO$_2$ (left) and MoS$_2$ on hBN (right), at 77 K. Insets shows the results at 300 K.